\documentclass[aps,prd,preprint,superscriptaddress,nofootinbib,amsfonts,amssymb,amsmath,longbibliography]{revtex4-1}
\usepackage{graphicx,bm,color,wrapfig}
\usepackage[colorlinks,linkcolor=red,anchorcolor=blue,citecolor=green]{hyperref}

\begin{document}

\title{Towards a unified interpretation of the early Universe in $R^2$-corrected dark energy model of $F(R)$ gravity}
\author{Hua Chen}
\email{huachen@mails.ccnu.edu.cn}
\affiliation{Institute of Astrophysics, Central China Normal University, Wuhan 430079, China}
\author{Taishi Katsuragawa}
\email{taishi@mail.ccnu.edu.cn}
\affiliation{Institute of Astrophysics, Central China Normal University, Wuhan 430079, China}
\author{Shinya Matsuzaki}
\email{synya@jlu.edu.cn}
\affiliation{Center for Theoretical Physics and College of Physics, Jilin University, Changchun, 130012, China}

\begin{abstract}
$R^2$-corrected dark energy (DE) models in $F(R)$ gravity have been widely investigated in recent years,
which not only removes the weak singularity potentially present in DE models
but also provide us with a unified picture of the cosmic history, including the inflationary and DE epochs. 
Towards the unified interpretation of dynamical DE all over the cosmic history in the class of $R^2$-corrected DE models, 
we explore the universal features of the scalaron dynamics in the radiation-dominated epoch,
along with the chameleon mechanism,
by keeping our eyes on the inflationary and DE epochs. 
We show that the scalaron evolution does not follow a \textit{surfing solution} and is mostly adiabatic before big bang nucleosynthesis (BBN), 
even properly including the \textit{kick} by the nonperturbative QCD phase transition, 
hence a catastrophic consequence claimed in the literature is not applied to this class of DE models. 
This is due to the presence of the gigantic scale hierarchy between $R^2$ correction and DE, so is the universal feature for the class of $R^2$-corrected DE models.  
The prospects for the post- or onset-inflationary epoch would be pretty different from what the standard $R^2$ inflationary scenario undergoes due to the presence of the chameleon mechanism. 
\end{abstract}

\maketitle

\section{Introduction}

Growing observational evidence indicates two accelerations of cosmic expansion:
the early-time inflation expected to occur before radiation domination~\cite{Planck:2018jri, Martin:2015dha},
and the late-time acceleration sourced by dark energy (DE)~\cite{SupernovaSearchTeam:1998fmf, SupernovaCosmologyProject:1998vns} (See~\cite{ParticleDataGroup:2020ssz} for reviews). 
There are two ways to realize an accelerated phase: to modify the matter content by introducing a new component, with the equation of state $w<-1/3$,
into the matter Lagrangian, or to modify the gravitational theory~\cite{Nojiri:2017ncd, Heisenberg:2018vsk}. 
For instance, $F(R)$ gravity is one of the simplest extensions of general relativity (GR) by improving the Ricci scalar $R$ to be an arbitrary function $F(R)$~\cite{DeFelice:2010aj}.
As a result, $F(R)$ gravity introduces one dynamical-field degree of freedom, dubbed \textit{scalaron},
which has the gravitational origin and could source the inflation and DE without requiring exotic components into the matter Lagrangian. 

Although phenomenologically successful,
it is known that viable DE models in $F(R)$ gravity generally suffer from a problem of the weak curvature singularity
with $\left|R\right|\to\infty$ at finite field value~\cite{Appleby:2008tv, Frolov:2008uf}.
A promising way to cure this problem is to add higher-curvature corrections,
for instance, the $\alpha R^2$ term~\cite{Dev:2008rx, Kobayashi:2008wc, Appleby:2009uf, Lee:2012dk}.
Moreover, $R^2$ gravity is also well known as the inflation model~\cite{Starobinsky:1980te},
which has proven very successful in light of the cosmic microwave background observation~\cite{Planck:2018jri}.
The two successes above may imply $R^2$ gravity as a viable {\it unified} theory. In addition to removing the curvature singularity,
$R^2$-corrected DE models could naturally realize the two accelerations by the same scalaron field~\cite{Odintsov:2016plw, Odintsov:2017hbk, Nojiri:2019fft, Oikonomou:2020qah,Odintsov:2019evb,Odintsov:2020nwm,Odintsov:2021wjz,Oikonomou:2022wuk}.
Other setups have also been considered, for instance, $R+\alpha R^{n} -\beta R^{2-n}$~\cite{Artymowski:2014gea} and its constraints~\cite{Geng:2015vsa, Yashiki:2020naf},
and a single scalar field extension to $F(R)$ gravity~\cite{Oikonomou:2020oex,Oikonomou:2022bqb}.

One may then wonder if such a unified model of $F(R)$ gravity can reproduce the overall cosmic history as much as we currently know. 
Although the fact of a gigantic hierarchy between the inflation and DE scales has allowed us to ignore one of two epochs in studying one,
$R^2$-corrected DE models in $F(R)$ gravity should be able to explain the intermediate epochs in the early Universe, such as the radiation-dominated epoch.
The modification for DE might have nontrivial effects on the potential structure of the scalaron, even in the early Universe. 
In that case, the complete picture of the whole cosmic history should be examined along with the nontrivial contributions to a {\it link} bridging over the early and later epochs. 
The research on such a link will shed light on the particle cosmology inspired by modified gravity in the early Universe. 

This work makes the first step along this new avenue: we investigate the dynamic scalaron evolution in the radiation-dominated epoch, including dynamics of matter fields, in correlation with the inflation and DE epochs.  
The exponential coupling with matter fields makes the scalaron possess an environment-dependent mass,
so that it can pass gravitational constraints in the local and dense solar systems and act as DE at the sparse cosmic scale~\cite{DeFelice:2010aj}.
Due to such a \textit{chameleon mechanism}, the decoupling of SM particles from the thermal equilibrium in the early Universe has a significant impact on the dynamics of the chameleon field in terms of scalar-tensor theories or the scalaron in $F(R)$ gravity.
In particular, it was shown that there exists a novel solution, called the surfing solution,
so that chameleons would be kicked by matter fields through the chameleon mechanism,
and be surfed in a constant and over the $\mathrm{GeV^2}$ velocity,
resulting in nonadiabatic changes in the chameleon's effective mass and the explosive production of scalarons with trans-Planckian momenta~\cite{Erickcek:2013oma, Erickcek:2013dea}.

Besides the SM-particle ensemble kicks (SMPE kicks) due to the particle decoupling,
we show that there is another powerful kick due to the QCD phase transition,
which we call the QCDPT kick in this work. 
The QCDPT kick would exacerbate the catastrophic consequence reported in~\cite{Erickcek:2013oma, Erickcek:2013dea}. 
Using data on the temperature dependence of the equation of state observed as the crossover reported from 
the nonperturbative lattice QCD simulation~\cite{Bali:2014kia}, 
we convert it into the change in the energy-momentum tensor around the QCDPT epoch and show that the QCDPT kick is more significant on the scalaron evolution than those estimated from the conventional thermal decoupling of the SM particles based on the perfect fluid approximation. 
Nevertheless, in terms of the scalaron in $F(R)$ gravity, we find that the surfing solution is always absent,  
owing to the unique and weak matter coupling. 
Thus, the kicks will be harmless and merely stabilize the scalaron at its effective potential minimum. 

We further find that the scalaron evolution at the radiation-dominated epoch is mostly adiabatic before BBN, so the catastrophic consequence claimed in the literature is not applied to this class at all. 
The null of catastrophe is universal in this class of $R^2$-corrected DE models because it essentially stems from the hierarchy between the $R^2$ correction and DE scales. 
Thus, the $R^2$ correction is also a solution to the catastrophic consequence, as well as the curvature singularity in general DE models of $F(R)$ gravity. 
The prospects for the post- or onset-inflationary epoch would differ from the standard $R^2$ inflationary scenario, which will also be briefly addressed. 

This paper is organized as follows.
In Section II, we briefly review the $R^2$-corrected DE model of $F(R)$ gravity and formulate the scalaron field equation in the flat Friedmann–Lema\^{i}tre–Robertson–Walker (FLRW) spacetime.
In Section III, we investigate the chameleon mechanism in the radiation-dominant epoch and demonstrate the kick solutions by the SMPE and QCDPT kicks.
Moreover, we evaluate the time evolution of the scalaron mass to show that the $R^2$ term suppresses the nonadiabaticity.
In Section IV, we briefly discuss the prospects of the post- or onset-inflationary epoch characteristic of the $R^2$-corrected DE models. 

Throughout this paper, a ``dot" or ``prime" denotes a derivative with respect to cosmic time or the argument of a function, respectively.
Especially $\varphi^{\prime}=d\varphi/d\tilde{N}$, where $\tilde{N}$ is the number of e-folds with ``tilde" denoting a physical quantity in the Einstein frame.

\section{$F(R)$ Gravity Theory}

\subsection{Rudiments of $F(R)$ Cosmology}

We begin by introducing the action of $F(R)$ gravity, which is given by
\begin{align}
    S=\int d^{4}x\ \sqrt{-g}\left[\frac{1}{2\kappa^{2}}F(R)+{\cal L}_{\text{M}}\left(g^{\mu\nu},\Phi_{\text{M}}\right)\right]
    \,,
\end{align}
where $\kappa^{2}\equiv 8\pi G \equiv 1/M_{\text{Pl}}^{2}$,
$g$ is the determinant of Jordan-frame metric $g_{\mu\nu}$,
and $\mathcal{L}_{\text{M}}$ is the matter Lagrangian as a function of $g_{\mu\nu}$ and matter fields $\Phi_{\text{M}}$.
Varying the action with respect to $g_{\mu\nu}$ gives the field equation
\begin{align}
    F'(R)R_{\mu\nu}-\frac{1}{2}F(R)g_{\mu\nu}-\nabla_{\mu}\nabla_{\nu}F'(R)+g_{\mu\nu}\square F'(R)
    =\kappa^{2}T_{\mu\nu}^{(\text{M})}
    \,,
\end{align}
where $\square\equiv\nabla^{\mu}\nabla_{\mu}$.
One finds that a dynamical field degree of freedom shows up when taking the trace
\begin{align}
    3\square F'(R)+F'(R)R-2F(R) =\kappa^{2}T_{\ \mu}^{\mu(\text{M})}
    \,.
\end{align}
Under the Weyl transformation,
\begin{align}
    \tilde{g}_{\mu\nu}&=e^{2\sqrt{1/6}\kappa\varphi}g_{\mu\nu}\equiv F'(R)g_{\mu\nu}
    \label{Weyl transformation}
    \,,
\end{align}
this field $\varphi$, dubbed scalaron, appears explicitly in the Einstein-frame action,
\begin{align}
    \tilde{S}=\int d^{4}x\ \sqrt{-\tilde{g}}\left[\frac{1}{2\kappa^{2}}\tilde{R}-\frac{1}{2}\tilde{g}^{\mu\nu}(\partial_{\mu}\varphi)(\partial_{\nu}\varphi)-V(\varphi)+e^{-4\sqrt{1/6}\kappa\varphi}\mathcal{L}_{\text{M}}\right]
    \,,
    \label{Einstein-frame action}
\end{align}
where the potential $V(\varphi)$ is defined as
\begin{align}
    V(\varphi)&\equiv\frac{1}{2\kappa^{2}}\frac{RF'(R)-F(R)}{F^{\prime2}(R)}
    \,,
    \label{Scalaron-vare-potential}
\end{align}
with $R=R(\varphi)$ given by the Weyl transformation in Eq.~\eqref{Weyl transformation}.
Varying Eq.~\eqref{Einstein-frame action} with respect to the Einstein-frame metric $\tilde{g}_{\mu\nu}$ yields
\begin{align}
    \tilde{R}_{\mu\nu}-\frac{1}{2}\tilde{R}\tilde{g}_{\mu\nu}-\kappa^{2}\left[-\frac{1}{2}\tilde{g}_{\mu\nu}\tilde{g}^{\rho\sigma}(\partial_{\rho}\varphi)(\partial_{\sigma}\varphi)+(\partial_{\mu}\varphi)(\partial_{\nu}\varphi)-\tilde{g}_{\mu\nu}V(\varphi)\right]&=\kappa^{2}\tilde{T}_{\mu\nu}^{(\text{M})}
    \,.
    \label{field equation}
\end{align}

In flat FLRW spacetime with perfect fluid approximation, $T_{\ \mu}^{\mu(\text{M})} = -\rho_\text{M} + 3P_\text{M}$,
Eq.~\eqref{field equation} gives the Friedmann equation,
\begin{align}
    3\tilde{H}^{2}=\kappa^{2}\left[\frac{1}{2}\left(\frac{d\varphi}{d\tilde{t}}\right)^{2}+V(\varphi)+\tilde{\rho}_{\text{M}}\right]
    \,,
    \label{Friedmann}
\end{align}
and the acceleration equation,
\begin{align}
    \frac{\ddot{\tilde{a}}}{\tilde{a}}=&-\frac{\kappa^{2}}{6}\left[2\left(\frac{d\varphi}{d\tilde{t}}\right)^{2}-2V(\varphi)+\tilde{\rho}_{\text{M}}+3\tilde{P}_{\text{M}}\right]
    \,.
\end{align}
Here, $a$ is the scale factor and $H \equiv \dot{a}/a$ is the Hubble parameter.
Varing Eq.~\eqref{Einstein-frame action} with respect to $\varphi$, the equation of motion of the scalaron reads
\begin{align}
    \frac{d^{2}\varphi}{d\tilde{t}^{2}}+3\tilde{H}\frac{d\varphi}{d\tilde{t}}=-\frac{dV}{d\varphi}-\frac{\kappa}{\sqrt{6}}e^{-4\sqrt{1/6}\kappa\varphi}T_{\ \mu}^{\mu(\text{M})}
    \,,
    \label{eom}
\end{align}
from which we define the effective potential ignoring the $\varphi$-dependence in the matter Lagrangian
\begin{align}
    V_{\text{eff}}(\varphi)&=V(\varphi)-\frac{1}{4}e^{-4\sqrt{1/6}\kappa\varphi}T_{\ \mu}^{\mu(\text{M})}
    \,.
    \label{effective potential}
\end{align}
Then the mass of the scalaron is defined as
\begin{align}
    m_{\varphi}^{2}\equiv V_{\text{eff}}^{\prime\prime}(\varphi)=V^{\prime\prime}(\varphi)-\frac{2\kappa^{2}}{3}e^{-4\sqrt{1/6}\kappa\varphi}T_{\ \mu}^{\mu(\text{M})}
    \label{mass}
    \,.
\end{align}

For later convenience, we define some dimensionless quantities:
the density ratio of cold matter $\tilde{\rho}_\text{m}$ and radiation $\tilde{\rho}_\text{r}$,
i.e. $f_\text{m}\equiv\tilde{\rho}_\text{m}/\tilde{\rho}_\text{r}$,
the kick function 
$\Sigma\equiv(\tilde{\rho}_{\text{r}}-3\tilde{P}_{\text{r}})/\tilde{\rho}_{\text{r}}$,
the scalaron $\phi\equiv\kappa\varphi$,
and the number of e-folds $\tilde{N}\equiv\ln(\tilde{a}/\tilde{a}_{i})$.
Then, Eqs.~\eqref{Friedmann} and~\eqref{eom} can be recast as
\begin{align}
    \tilde{H}&=\kappa\sqrt{\frac{V+\left(1+f_{\text{m}}\right)\tilde{\rho}_{\text{r}}}{3\left(1-\frac{1}{6}\phi^{\prime2}\right)}}
    \,,
    \label{Friedmann2}
\end{align}
and
\begin{align}
    \frac{V+\left(1+f_{\text{m}}\right)\tilde{\rho}_{\text{r}}}{3\left(1-\frac{1}{6}\phi^{\prime2}\right)}\phi^{\prime\prime}+\left(V+\frac{2+3f_{\text{m}}+\Sigma}{6}\tilde{\rho}_{\text{r}}\right)\phi^{\prime}&=-\frac{d V}{d\phi}+\frac{1}{\sqrt{6}}\left(\Sigma + f_{\text{m}}\right)\tilde{\rho}_{\text{r}}
    \,,
    \label{eom2}
\end{align}
where $\phi^\prime=d\phi/d\tilde{N}$.
Since in the radiation dominated epoch, $f_\text{m}\lesssim 10^{-6}$~\cite{Erickcek:2013dea}, we shall hereafter ignore $f_\text{m}$.
Finally, the scalaron velocity with respect to the cosmic time can be easily read off as 
\begin{align}
    \frac{d\varphi}{d\tilde{t}}
    =\phi^{\prime}\sqrt{\frac{V+\tilde{\rho}_{\text{r}}}{1-\frac{1}{6}\phi^{\prime2}}}
    \label{velocity}
    \,.
\end{align}

\subsection{$R^2$-corrected DE models}

We now consider $R^2$-corrected DE models with  
\begin{align}
    F(R) = F_{\text{DE}}(R) +\alpha R^{2}
    \label{F(R)_unification_model}
    \,,
\end{align}
and take the Starobinsky's DE model as a benchmark:
\begin{align}
    F(R) 
    = R -\beta R_{\text{c}}\left[1-\left(1+\frac{R^{2}}{R_{\text{c}}^{2}}\right)^{-n}\right]
    +\alpha R^{2}
    \,,
    \label{Starobinsky DE}
\end{align}
where $\beta > 0$, $n > 0$, and $R_\text{c}\sim\Lambda\simeq4\times10^{-84}\ [\mathrm{GeV}^{2}]$ are model parameters.
The potential is a multi-valued function (see, e.g., Refs.~\cite{Frolov:2008uf, Katsuragawa:2019uto}). 
In the present study we focus on the case where $+\infty>R>R_\text{b}$ with the  branch located at $R_\text{b} \simeq 0.93 R_\text{c}$ for $\beta=2$ and $n=1$. The fifth force experiment constrains $\alpha<10^{22}\ [\mathrm{GeV}^{-2}]$~\cite{Cembranos:2008gj},
and $\alpha = 1/(6 m_0^2) \sim 10^{-27}\ [\mathrm{GeV}^{-2}]$ if one improves $\alpha R^2$ to be responsible for the inflation with mass scale $m_0 \sim 10^{13}\ \mathrm{[GeV]}$.

\begin{figure}[htbp]
    \centering
    \includegraphics[scale=0.5]{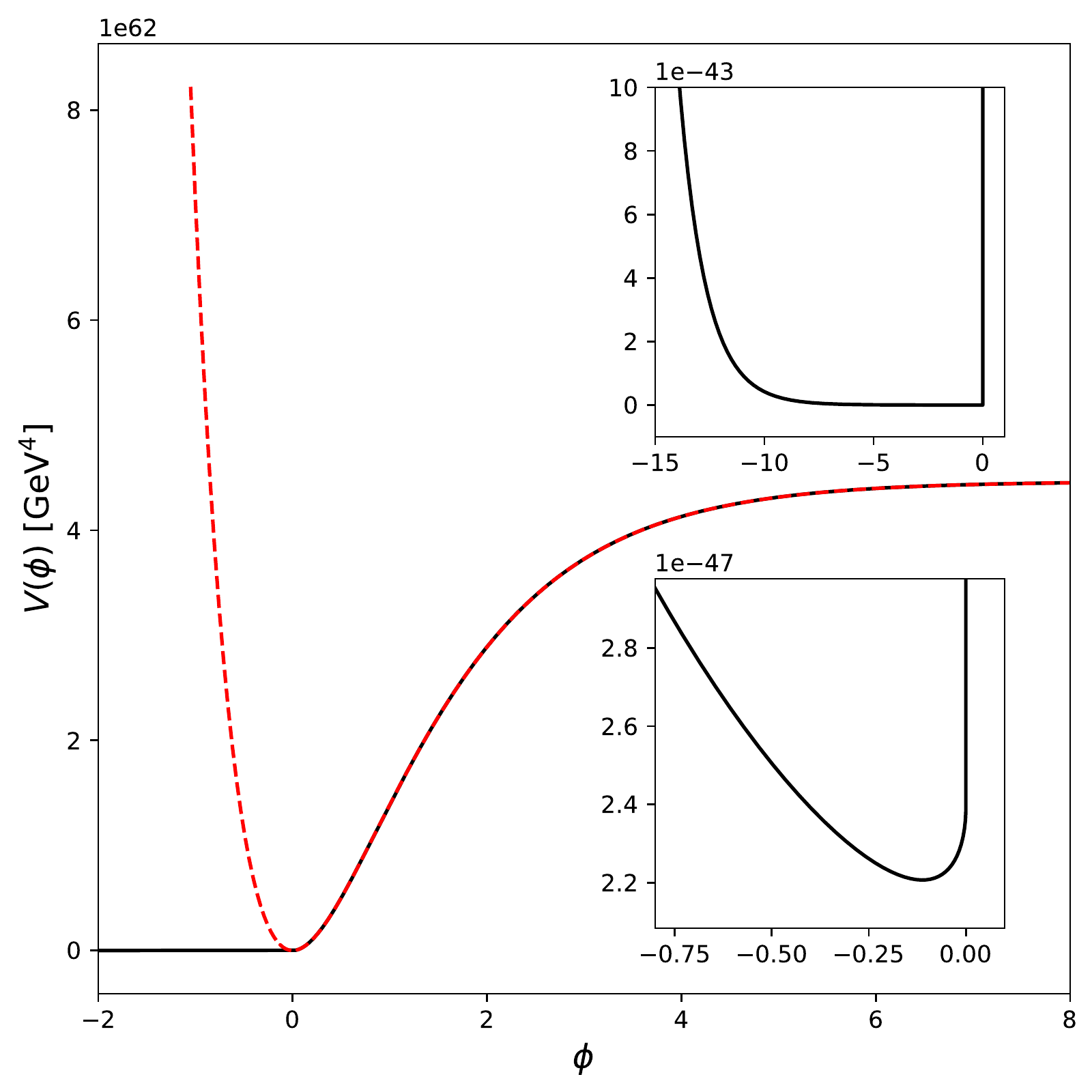}
    \caption{
        Black solid curve shows the potential of the scalaron in the class of $R^2$-corrected model for $F(R)$ gravity, where we have chosen $\alpha = 1/6\times 10^{-26}\ \mathrm{[GeV^{-2}]}$, $\beta=2$ and $n=1$. 
        The scalaron field has been normalized by the Planck scale in the plot ($\phi = \kappa \varphi$). 
        The potential minimum is set so as to realize the DE scale
        $V_{0}\equiv\frac{R_{\text{c}}}{2\kappa^{2}}\sim\rho_{\Lambda}\approx10^{-47}\ [\mathrm{GeV}^{4}]$. 
        Though being visibly shallow at the potential minimum, the potential asymptotically reaches infinity, $V(\varphi) \to \infty$, when $\varphi \to -\infty$.
        The red dashed curve shows $R^{2}$ inflation for comparison.
        }
    \label{Fig: bare_potential}
\end{figure}

As is in the case of the $R^{2}$ inflation, 
the scalaron potential has a flat plateau, which realizes the slow-roll inflation,
as shown in Fig.~\ref{Fig: bare_potential}.
On the contrary to $R^{2}$ inflation,
the potential possesses a shallow plain with a global minimum at $\varphi \simeq 0.11 M_\text{Pl}$,
responsible for the DE vacuum in the current Universe.
Therefore, one can ignore the DE modification only when discussing the inflationary epoch triggered by the $R^{2}$ term;
as will be clarified later, however, the DE modification can alter the standard cosmic history after the inflation, even in the early Universe.

We note that the structure of the potential above is common among the unified models of the $R^{2}$ inflation and DE. 
There should be two (almost) flat parts of potential to realize the two accelerations in the early and late-time Universe, 
because cosmological observations indicate that inflation is described by quasi-de Sitter spacetime in the early Universe and that DE almost acts as the cosmological constant in the late-time Universe. 
Since the viable DE model should reproduce the $\Lambda$-CDM model in the large-curvature limit,
the model-dependence is negligible in the regime where $\phi>0$.
Moreover, considering the matter contribution to the effective potential, the flat potential for the DE background is overwhelmed by the trace of the energy-momentum tensor in the region where $\phi<0$.
In other words, we can discuss the intermediate stage between the above two accelerations in a model-independent way,
although our analysis in the present work specifies the DE model to the Starobinsky model.

Furthermore, the potential structure connecting two flat potentials by a gigantic scale difference is related to the hierarchy problem of DE. 
Thus, even though the $R^{2}$ correction will not be identified as $R^{2}$ inflation, 
satisfying the constraint on the parameter $\alpha$, 
the flat potential for the DE background is always smaller than that for the $R^{2}$ correction.
The details of potential analysis and argument about the model independence are summarized in Appendix.~\ref{appendix A}.

\section{Scalaron in Radiation domination}

\subsection{Absence of Surfing Solution}

After the production of SM particles via the reheating epoch, 
the Universe is almost full of radiation,
with the trace of the energy-momentum tensor vanishing.
Due to the cooling of the Universe, however,
SM particles decouple successively from the thermal plasma,
resulting in nonzero contributions to the trace of the energy-momentum tensor,
and thereby affect the scalaron dynamics due to the matter coupling.
Therefore, first of all, we estimate the energy density at the radiation-dominated epoch, 
$\rho_{\text{r}}(T)$, 
and $\Sigma (T)$. 

Assuming the conventional ensemble of SM particles in the perfect fluid 
to well describe the early Universe, 
the trace of the energy-momentum tensor for each relativistic particle is evaluated 
\begin{align}
    \rho_{i}-3P_{i}&=\frac{g_{i}T^{4}}{2\pi^{2}}x^{2}\int_{0}^{\infty}dy\ \frac{y^{2}}{\sqrt{x^{2}+y^{2}}}\frac{1}{\exp\left(\sqrt{x^{2}+y^{2}}\right)\pm1}
    \label{fluid_trace_EMT}
    \,,
\end{align}
where $g_{i}$ is the degree of freedom, $x\equiv m_{i/T}$, $y\equiv P_{i}/T$,
with ``$i$'' labelling different SM species,
and ``$+/-$'' are applied to fermion and boson, respectively.
The energy density of thermal bath $\rho_\text{r}$ is obtained by
\begin{align}
    \rho_\text{r}\equiv\frac{\pi^2}{30}g_{*}(T)T^4
    \,,
    \label{energy density}
\end{align}
where $g_{*}(T)$ is relativistic effective degrees of freedom.

Then the kick function $\Sigma(T)$ reads
\begin{align}
    \Sigma_{i}(T)
    =\frac{\rho_{i}-3P_{i}}{\rho_{\text{r}}}=\frac{15}{\pi^{4}}\frac{g_{i}}{g_{*}(T)}x^{2}\int_{0}^{\infty}dy\frac{y^{2}}{\sqrt{x^{2}+y^{2}}}\frac{1}{\exp\left(\sqrt{x^{2}+y^{2}}\right)\pm1}
    \label{kick function}
    \,.
\end{align}
Summing up all contributions from the thermal decoupling of SM particles,
we find four peaks corresponding to what we call the SMPE kicks,
which would affect the scalaron evolution significantly, as will be seen later. 

Using the conservation of the entropy density per comoving volume in the Jordan frame, 
$sa^{3}=\frac{2\pi^{2}}{45}g_{*s}(T)T^{3}a^{3}=\text{const.}$, 
one can extract $\Sigma (\tilde{N})$
by making a replacement 
\begin{align}
    Tg_{*s}^{1/3}(T)=g_{*s}^{1/3}(T_{i})T_{i}e^{\sqrt{1/6}\left(\phi-\phi_{i}\right)-\tilde{N}}
    \label{entropy conservation}
\end{align}
to obtain $T(\tilde{N})$.

It has been shown that there exists a novel solution, dubbed surfing solution, in scalar-tensor theories,
which would cause nonadiabaticity and explosive particle production~\cite{Erickcek:2013oma, Erickcek:2013dea}.
Here, however, we show that {\it in the case of $F(R)$ gravity, 
there exists no surfing solution.} 

In scalar-tensor theories, the conformal factor, related by 
the Weyl transformation, as in Eq.(\ref{Weyl transformation}), 
includes free parameters dependent of particle spices, $Q_i$, like
\begin{align}
    \tilde{g}_{\mu\nu}=e^{2Q_i\kappa\varphi}g_{\mu\nu}
    \,. 
\end{align}
In the case of $F(R)$ gravity,
we have the universal coupling $Q=Q_i =1/\sqrt{6}$. 
Since the DE scale is very small, one can assume that
$V(\phi)\ll\rho_\text{r}$ and $dV/d\phi$ is negligible
compared to the driving term in $\Sigma$ in the early Universe. 
Then Eq.~\eqref{eom2} is reduced to
\begin{align}
    \phi^{\prime\prime}
    =\left(1-\frac{1}{6}\phi^{\prime2}\right)\left[-\left(1+\frac{\Sigma}{2}\right)\phi^{\prime}+3Q\Sigma\right]
    \,.
\end{align}
Eq.~\eqref{entropy conservation} allows for the existence of a surfing solution,
i.e. $\phi^{\prime\prime}=0$ and $\phi^{\prime}=1/Q$.
This solution is valid if and only if the kick function satisfies
\begin{align}
    \Sigma(T_{\text{s}})\geq\frac{2}{6Q^{2}-1}
    \,.
    \label{surfing condition}
\end{align}
One finds that this condition does not hold in $F(R)$ gravity with $Q=1/\sqrt{6}$.  
Thus, the surfing solution is absent in $F(R)$ gravity no matter how the kick function varies.
Note that our condition is different from that in~\cite{Erickcek:2013oma, Erickcek:2013dea}
since we do not assume that $1+\Sigma\simeq 1$ in this work.
We also emphasize that the absence of the surfing solution is independent of specific $F(R)$ models
because the consequence relies only on the coupling constant $Q_{i}$ uniquely determined by the Weyl transformation,
although it is an arbitrary parameter in the general scalar-tensor theory.

\subsection{SMPE and QCDPT kicks}

\begin{figure}[htbp]
    \centering
    \includegraphics[scale=0.6]{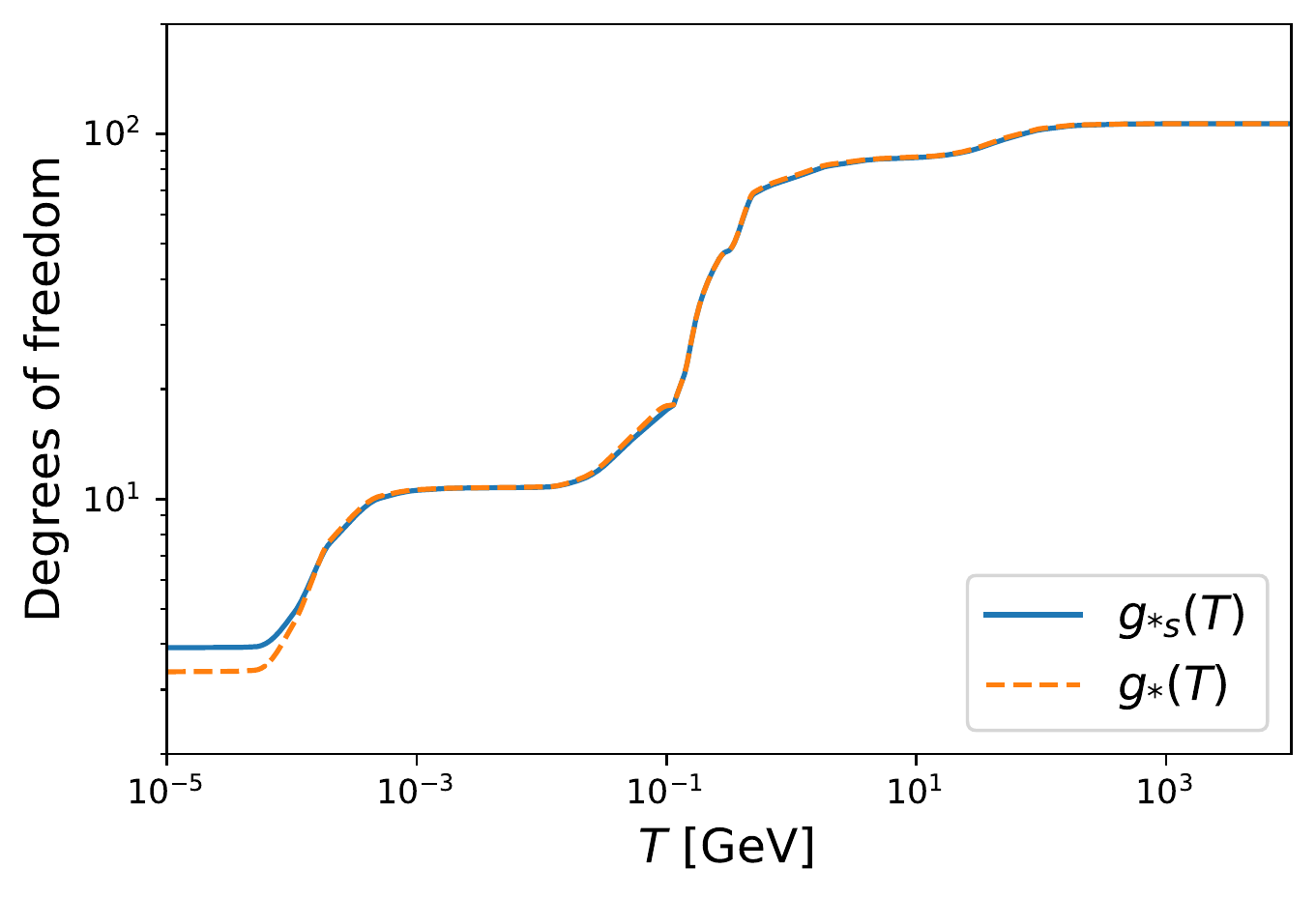}
    \caption{
        The effective degrees of freedom as a function of the Jordan-frame temperature for entropy (solid blue line) and energy density (orange dashed line).
        }
    \label{Fig: dof}
\end{figure}

\begin{figure}[htbp]
    \centering
    \includegraphics[scale=0.55]{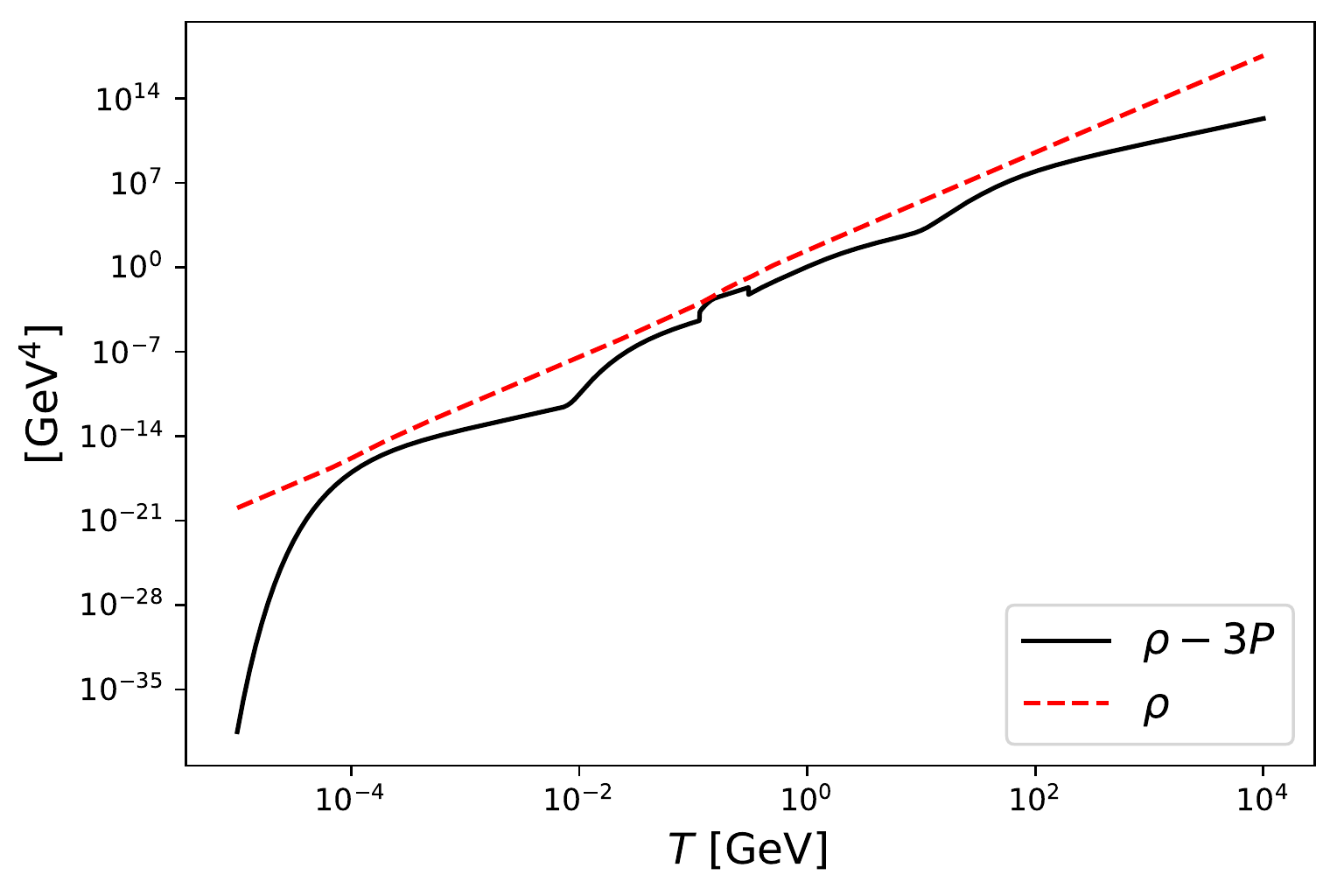}
    \caption{
        Comparison of the trace of the energy-momentum tensor and the energy density of radiation as a function of the Jordan-frame temperature.
        }
    \label{Fig: trace_rho}
\end{figure}

\begin{figure}[htbp]
    \centering
    \includegraphics[scale=0.55]{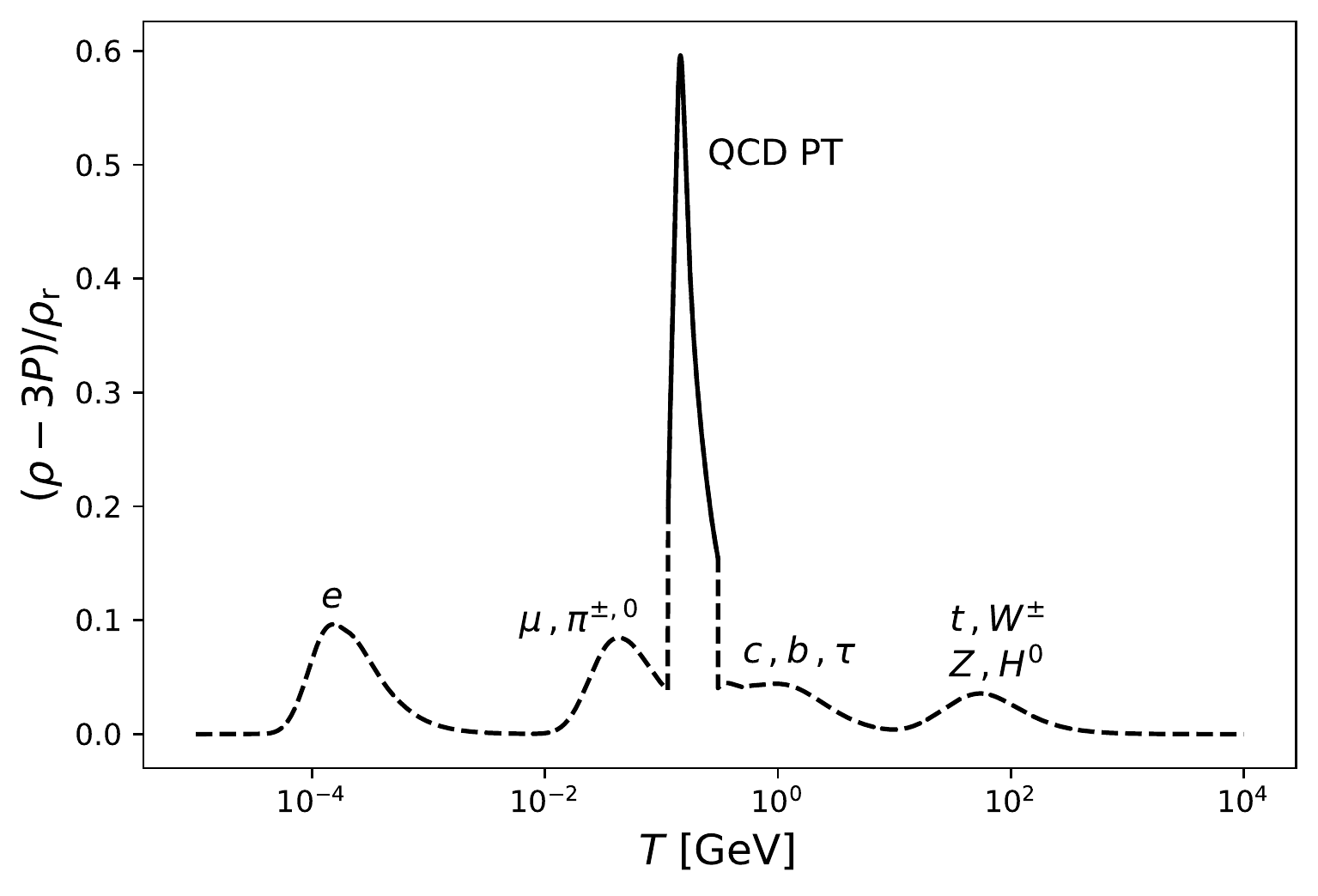}
    \caption{
        The kicks as a function of the Jordan-frame temperature during the radiation-dominant epoch, including the QCD phase transition and the perfect-fluid description ensemble SM particles in the thermal equilibrium. 
        The solid curve around the center of the plot denotes the contribution from the QCDPT kick.
        The dashed curves have been drawn by accumulating the ensemble contributions from the thermal decoupling of SM particles.
        }
    \label{Fig: kicks}
\end{figure}

It turns out that even without a surfing solution, the SMPE and QCDPT kicks have a nontrivial impact on the dynamics of scalaron.
Utilizing data about effective degrees of freedom from Ref.~\cite{Husdal:2016haj} together with the result on the equation of state from the lattice QCD simulation~\cite{Bali:2014kia},
ranging from $T=114-305$~[MeV],
we plot the effective degrees of freedom for the energy density and entropy density in Fig.~\ref{Fig: dof},
compare the trace of the energy-momentum tensor with the energy density in Fig.~\ref{Fig: trace_rho},
and thereby obtain the kick function in Fig.~\ref{Fig: kicks}.
We find that the strongest kick among the SMPE kicks comes from the electron-positron annihilation, which yields $\Sigma \simeq 0.1$,
while the QCDPT kick is by six times greater: $\Sigma \simeq 0.6$.
The detail of data handling is shown in Appendix.~\ref{appendix B}.

\begin{figure}[htbp]
    \centering
    \includegraphics[scale=0.6]{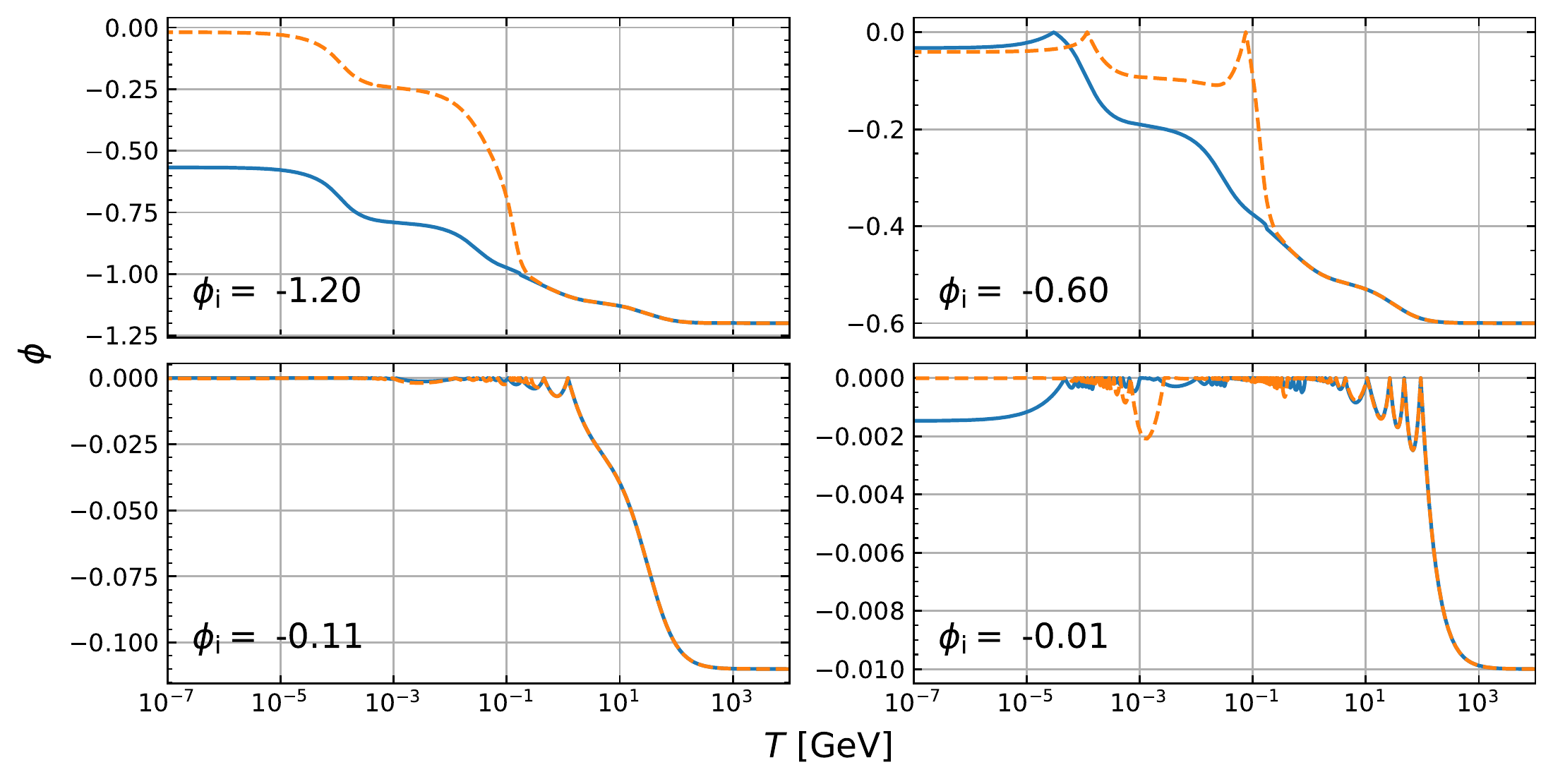}
    \caption{
        The scalaron evolution under the SMPE kicks (blue solid line) and total kicks (orange dashed line),
        where just as sample points, we have chosen initial conditions
        $\varphi_i/M_\text{Pl} = - (1.2, 0.6, 0.11, 0.01)$ at 
        $T_{i} = 10000$ [GeV],
        with $g_{*s}\left(T_{i}\right) = 106.75$.
        }
    \label{Fig: scalaron}
\end{figure}

We discuss the implications of the QCDPT kick to the scalaron evolution, 
in comparison with the four conventional SMPE kicks. 
First of all, we note that due to the big hierarchy of the inflation and DE scales,
the scalaron cannot climb up the potential barrier at $\varphi \geq 0$,
so that it rebounds after hitting the potential barrier,
as seen from Fig.~\ref{Fig: scalaron}. 
Meanwhile, we observe that the four SMPE kicks and full five kicks push the scalaron \textit{at most}
$0.64\ M_\text{Pl}$ and $1.41\ M_\text{Pl}$, respectively,
which means that the QCDPT kick is more effective than the SMPE kicks.
In terms of scalar-tensor theories, the surfing solution takes effect if $Q\geq\sqrt{\frac{2+\Sigma(T_{\text{s}})}{6 \Sigma(T_{\text{s}})}}\gtrsim 0.85$,
and the velocity can be as large as $\phi'=1/Q\gtrsim 1.17$, 
which exacerbates the existing catastrophic results claimed in~\cite{Erickcek:2013oma, Erickcek:2013dea}. 

On the other hand, as seen from Fig.~\ref{Fig: scalaron}, the matter contribution related to the chameleon mechanism in $F(R)$ gravity gets the opposite effect compared with that in scalar-tensor theories.
The chameleon mechanism indeed protects the scalaron from leaving the potential minimum.
In other words, the balance between the kicks, bare potential, and Hubble friction will eventually stabilize the scalaron at the potential minimum.
This brings two advantages to $F(R)$ gravity compared with other models:
1. surfing solution is absent in $F(R)$ gravity; 2. No specific initial conditions are required.

To examine the second point, we determine the scalaron field value in the BBN epoch.
First, note that the Weyl transformation, in Eq.~\eqref{Weyl transformation}, implies that a constant mass scale $m_{i}$ in the Jordan frame will be modified due to the $\varphi$ dependence like $\tilde{m}_i=m_{i}\exp\left(-\sqrt{1/6}\kappa\varphi\right)$ in the Einstein frame. 
Thus the proton-neutron mass difference could also be altered by 
the presence of the scalaron field with somewhat a large field value. 
As was addressed in~\cite{Chen:2019kcu}, 
the size of the scalaron filed, normalized to the Planck scale,
is constrained so as not to spoil the successful BBN. 
When a conservative limit of 10\% in size (in the unit of the Planck scale) is taken,
we find that the allowed interval of the scalaron field value during the BBN epoch is $-0.33\lesssim\kappa\varphi_{\text{BBN}}\lesssim 0.15$,
which requires the initial value as $\varphi_i\gtrsim -0.97\ M_\text{Pl}$
or $\varphi_{i}\gtrsim -1.74\ M_\text{Pl}$ under the SMPE kicks or the total five kicks. 
More stringent bound as evaluated in~\cite{Chen:2019kcu} might be placed,  
where $-0.05\lesssim\kappa\varphi_{\text{BBN}}\lesssim0.03$. 
In that case, we would have $\varphi_i\gtrsim -0.69\ M_\text{Pl}$ with only the four SMPE kicks taken into account, 
or $\varphi_{i}\gtrsim -1.46\ M_\text{Pl}$ with the total kicks considered\footnote{
We have assumed the field value to be negative at this point since we just regard the unified model as a dark energy model and have not related it with inflation yet.}.
Thus, the QCDPT kick is also significant in relaxing the initial value of the scalaron field.

\section{Nonadiabaticity in Dynamics of Scalaron}

\begin{figure}[htbp]
    \centering
    \includegraphics[scale=0.6]{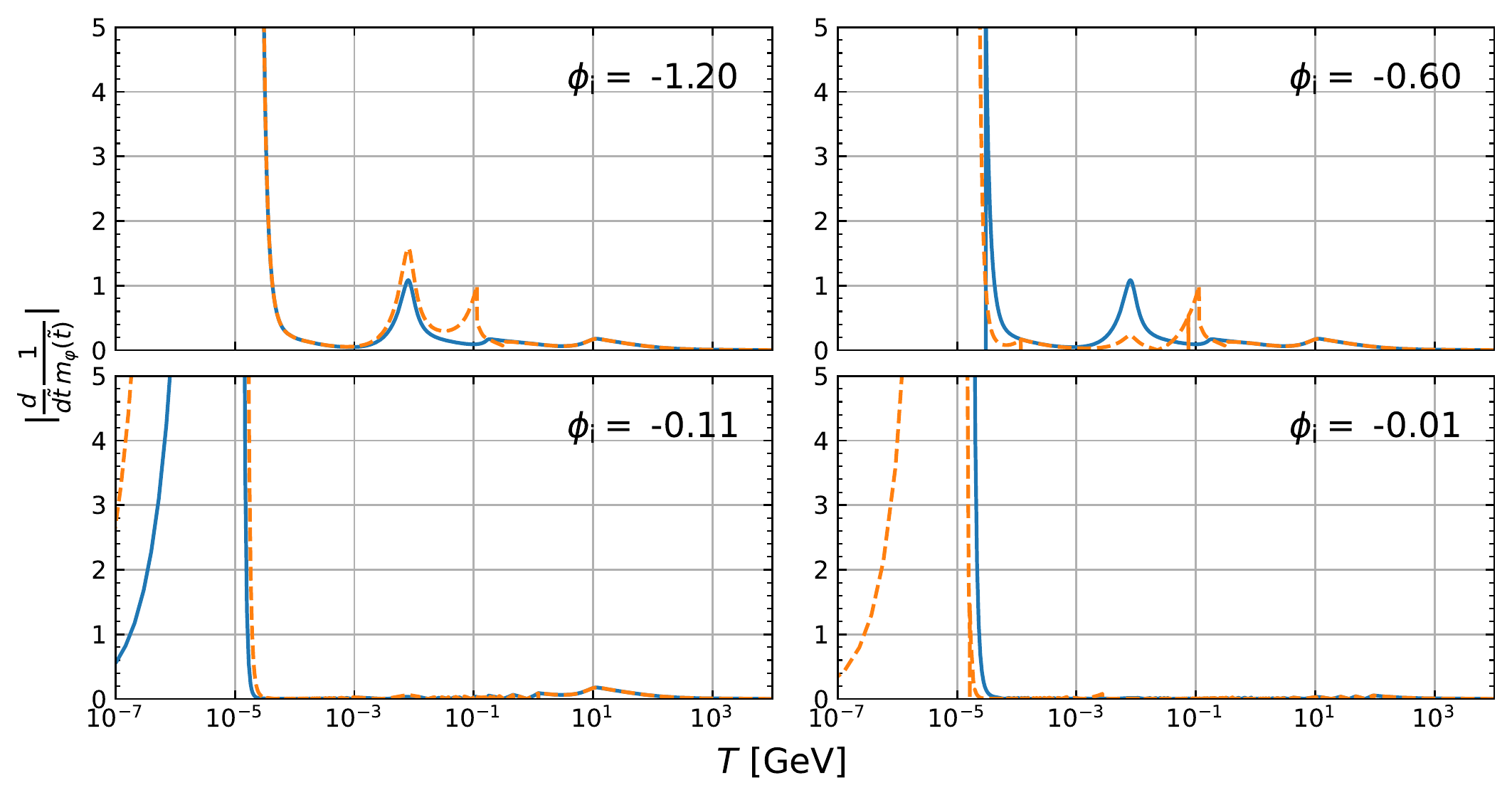}
    \caption{
        The nonadiabaticity as a function of the Jordan-frame temperature,
        where the solid-blue line and orange dashed line denote that the scalaron is under SMPE kicks and total kicks, respectively.
        The initial conditions are chosen as the same as in Fig.~\ref{Fig: scalaron}.
        }
    \label{Fig: scalaron_nonadiabaticity}
\end{figure}

The DE scale is extremely small compared to other characteristic scales in the early Universe,
in which scale distance the scalaron travels. Hence the motion of the scalaron could get too fast to be adiabatic.
The adiabaticity is violated when 
\begin{align}
    \left|\frac{d}{d\tilde{t}}\frac{1}{m_{\varphi}\left(\tilde{t}\right)}\right|\gtrsim 1
    \,.
\end{align}
From Eqs.~\eqref{mass} and~\eqref{velocity}, we have
\begin{align}
    \left|\frac{d}{d\tilde{t}}\frac{1}{m_{\varphi}\left(\tilde{t}\right)}\right|
    &=\left|-\frac{\phi^{\prime}}{2}\sqrt{\frac{V(\phi)+\tilde{\rho}_{\text{r}}}{3\left(1-\frac{1}{6}\phi^{\prime2}\right)}}\frac{V^{\prime\prime\prime}(\phi)-\frac{8}{3\sqrt{6}}\Sigma\tilde{\rho}_{\text{r}}}{\left(V^{\prime\prime}(\phi)+\frac{2}{3}\Sigma\tilde{\rho}_{\text{r}}\right)^{3/2}}\right|
    \,.
    \label{na_kicks}
\end{align}
In Fig.~\ref{Fig: scalaron_nonadiabaticity} we plot this discriminator of the adiabaticity as a function of the Jorndan-frame temperature in the radiation dominated epoch.  

As shown in Fig.~\ref{Fig: scalaron_nonadiabaticity}, the strong signals of the nonadiabaticity appear just after the electron-positron annihilation at $T\sim 10^{-5}~[\mathrm{GeV}]$. 
Although the results depend on the initial condition we have chosen,
we numerically confirmed that the large violation of adiabaticity below $T=10^{-5}~[\mathrm{GeV}]$ is insensitive to the DE potential and thus the energy density of matter.
Neglecting the DE scale in Eq.~\eqref{na_kicks}, the discriminator is approximated to be 
\begin{align}
    \left|\frac{d}{d\tilde{t}}\frac{1}{m_{\varphi}\left(\tilde{t}\right)}\right|
    \approx\left|\phi^{\prime}\sqrt{\frac{1}{3\Sigma\left(1-\frac{1}{6}\phi^{\prime2}\right)}}\right|
    \label{na_kicks_approx}
    \,.
\end{align}
From this form, we see that since in the conventional perfect fluid description, 
the kick function $\Sigma(T)$ goes vanishing after the electron-positron annihilation 
as shown in Fig.~\ref{Fig: kicks},
it is unavoidable that the adiabaticity is badly violated.
The rapid drop of the kick function is mainly due to the fast decrease of the energy density of radiation,
compared with the trace of the energy-momentum tensor,
as shown in Fig.~\ref{Fig: trace_rho}.
One also notes that the adiabatic condition is slightly violated at around $T=10^{-2}$~[GeV], as shown in the upper panel in Fig.~\ref{Fig: scalaron_nonadiabaticity}. Again, it is due to the faster drop of the trace of the energy-momentum tensor than the energy density. 
Those rapid drops might be smeared to be not so much sharper than what we observe in Fig.~\ref{Fig: kicks},
by taking into account not only the conventional ensemble of single free quanta but also interacting multi-body processes (as encoded in collision terms in the Boltzmann equation) in evaluating $\Sigma$.  
Therefore, we claim that emergence of the nonadiabaticity could be an artifact coming from the invalidity of the conventional ensemble description. 

We could qualitatively argue the validity of the conventional ensemble description applied in the existing works.
First of all, note that the matter contents can be categorized into two, i.e., relativistic radiation and nonrelativistic matter. 
When the temperature $T$ is lower than the mass $m$, 
Eq.~\eqref{fluid_trace_EMT} suggests $\rho_{i} - 3P_{i} \approx \rho \propto T^{3/2} e^{-m/T}$.
Since $\rho_\text{r} \propto T^{4}$, 
the kick function Eq.~\eqref{kick function} in the thermal equilibrium evolves as
\begin{align}
    \Sigma(T) \propto T^{-5/2} e^{-m/T}
    \,.
\end{align}
Thus, as the temperature decreases, the kick function exponentially decreases to zero,
which causes the strong nonadiabaticity in Eq.~\eqref{na_kicks_approx}.

On the other hand, the above picture is invalid after the thermal decoupling of SM particles, and the number density $n$ satisfies $n \propto a^{-3}$ to produce the thermal relic abundance.
Because the scale factor and temperature are related to each other as $a \propto T^{-1}$, 
the energy density should follow $\rho = m n \propto T^{3}$ after the decoupling.
Thus, taking into account the effect of decoupling, we find
\begin{align}
    \Sigma(T) \propto T^{-1}
    \,.
\end{align}
Therefore, the kick function no longer decreases monotonically, which can prevent the scalaron from being nonadiabatic.
It is remarkable that the decoupling significantly affects the scalaron dynamics in the thermal history,
and it shows an adaptive limit of the conventional ensemble description in the study of the chameleon mechanism.
Note that the chameleon mechanism works to stabilize the scalaron with the thermal relics of SM particles and to avoid the nonadiabaticity.

\section{Conclusion and Discussions}

The unification of DE and inflation in $F(R)$ gravity has been widely investigated.
This work has explored the scalaron evolution with a specific potential structure combining the DE and inflationary epochs.
The scalaron should have been frozen around its de Sitter vacuum in the radiation-dominated epoch due to Hubble friction. 
However, as shown in this work, the decoupling of SM particles from the thermal bath would drive the scalaron to evolve and climb its effective potential.
In particular, the contribution from the QCD PT has a nontrivial impact on the scalaron dynamics.
We found that the kicks with/without the QCD PT would pull the scalaron up to $0.64\ M_\text{Pl}$ and $1.41\ M_\text{Pl}$, respectively.

As a result, 
successful BBN is preserved as long as one chooses the proper initial conditions of the scalaron field at the beginning of the radiation-dominated epoch.
Unlike usual chameleon models, due to the scale hierarchy of inflation and DE, 
we clarified that the $R^2$ term protects the scalaron mass from receiving nonadiabatic change in time.
On the other hand, the nonadiabaticity relies on the choice of the initial value,
which is determined by the inflationary epoch. 
Therefore, to estimate the impact of the plain part of the potential on the scalaron dynamics,
one has to study the inflation, reheating, and radiation-dominated epochs systematically and inclusively.
Studying such intermediate stages is crucial in assessing the validity of $F(R)$ gravity,
and we will leave the complete analysis in future work. 

In closing, we make possible prospects for the post/onset-inflationary epoch and particle production to supply the reheating of the Universe.
In the current work, we have discussed the impact of the $R^2$ term at the radiation-dominated epoch,
whose potential is so steep that the scalaron hardly climbs up even under the most potent QCDPT kick.
Although we demonstrated the numerical analysis for four different initial conditions chosen by hand,
the ambiguity of the initial condition can be embedded into the inflation dynamics if we consider the whole cosmic history.
Thus, it is inevitable to investigate the scalaron dynamics between the inflation and radiation-dominated epoch.
We shall briefly discuss the connection of $R^2$ inflation to DE.
Although the DE scale has little impact on the dynamics of inflation~\cite{Geng:2015vsa},
the post inflationary dynamics, i.e. (p)reheating, would be significantly changed due to the presence of plain at $\varphi<0$.

\begin{figure}[htbp]
    \centering
    \includegraphics[scale=0.6]{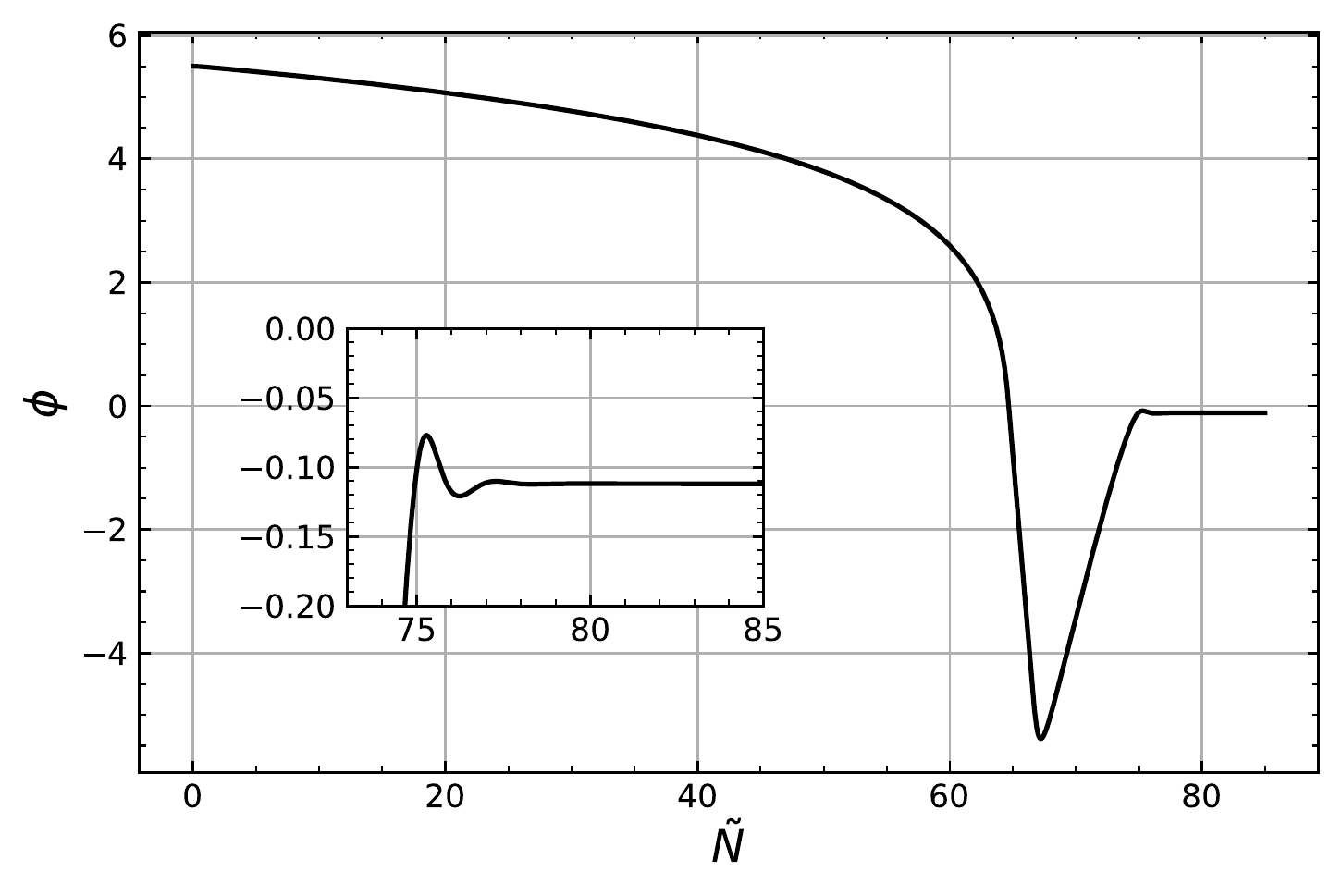}
    \caption{
        The scalaron evolution in absence of matter fields, where $\tilde{N}$ is the number of e-folds in the Einstein frame. In order for enough e-folds during inflation, we have chosen $\phi_\text{i}=5.5$.}
    \label{Fig: inflation}
\end{figure}

\begin{figure}[htbp]
    \centering
    \includegraphics[scale=0.6]{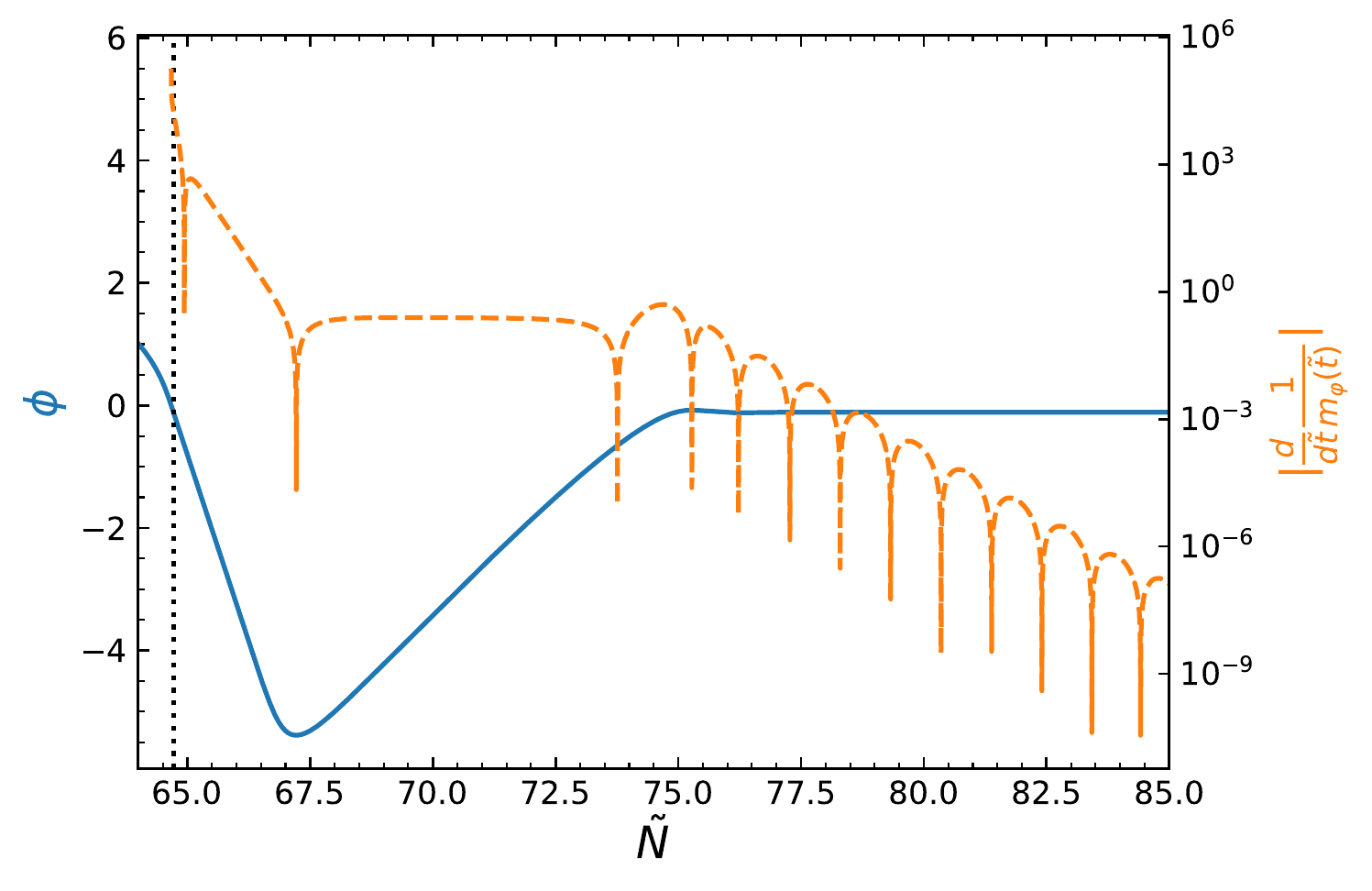}
    \caption{
        The nonadiabaticity of the scalaron in the absence of matter fields,
depicted by the orange-dashed line.
        The blue-solid line corresponds to the field evolution (the same as Fig.~\ref{Fig: inflation}),
        while the black-dotted line marks the corresponding e-folding number when the scalaron reaches the bare potential minimum.
        }
    \label{Fig: na_inf}
\end{figure}

Assuming the cold inflation, the reheating after the inflation induces the radiation-dominated epoch.
In absence of matter fields during the inflation, one observes that the equation of motion is reduced to
\begin{align}
    \phi^{\prime\prime} = -3\left(1-\frac{1}{6}\phi^{\prime2}\right)\left(\phi^{\prime}+\frac{dV}{Vd\phi}\right)
    \,.
\end{align}
By solving the equation of motion,
as shown in Fig~\ref{Fig: inflation},
we find the scalaron would quickly pass its potential minimum at $\varphi\simeq -0.11M_\text{Pl}$,
experiences one single rebound, and eventually oscillates at around the effective potential minimum.
We also reevaluate the nonadiabaticity in the scalaron mass in Eq.~\eqref{na_kicks} with the matter contribution ignored. 
As in Fig.~\ref{Fig: na_inf}, we find that the nonadiabaticity is triggered when the scalaron passes through its bare potential minimum.
Note that $V^{\prime \prime}(\varphi)$ is negative during the slow-roll inflation, and we cannot define the mass in Eq.~\eqref{na_kicks}. 

The strong nonadiabaticity around the bare potential minimum originates from the hierarchy between the inflation and DE scales:
that is,  the role-shift of the scalaron from the inflaton to dynamical DE triggers the huge change in time for the scalaron mass from the inflationary epoch to the post-inflationary epoch.
The nonadiabaticity indicates the nonperturbative self-production of the scalaron quanta and the breakdown of classical analysis~\cite{Erickcek:2013oma, Erickcek:2013dea}.
Consequently, compared with the original $R^{2}$ inflation, the perturbative oscillation picture around $\varphi=0$ disappears, 
thus, in the unified $R^2$-corrected class of models in $F(R)$ gravity, {\it we need to alter the post-inflationary dynamics of the standard cosmology}.
This consequence relies only on the hierarchy between the inflation and DE scales, 
irrespective of the details of DE models~\cite{Motohashi:2012tt,Nishizawa:2014zra}.

Therefore, refining the early cosmology in $R^2$-corrected DE models of $F(R)$ gravity is mandatory.
With the results based on the current study, the following approaches would be plausible:
\begin{itemize}
\item 
In light of the nonadiabaticity of the scalaron, 
the effective scalaron mass would presumably include correlation functions of the scalaron quanta~\cite{Erickcek:2013oma, Erickcek:2013dea}.
As in Fig.~\ref{Fig: na_inf}, the scalaron self-production due to the large nonadiabaticity could happen when the scalaron rolls the potential, but before it passes through the DE vacuum and rebounds ($\tilde{N} \lesssim 67.5$).
Then, the produced and accumulated scalaron quanta can highly dominate in the effective potential over the bare potential 
of the DE scale.
Then it acts as the potential barrier to push the scalaron back to $\varphi=0$.
In other words, the energy density of the scalaron quanta serves as the chameleon mechanism. It lifts the potential, which prevents the scalaron from going to the bare potential minimum. 
This barrier would instantaneously be present and gone when the nonperturbative-scalaron quanta-production ends. 
Then the background scalaron could get escaped back to the standard DE vacuum. 
Such a {\it self-chameleon mechanism} could trigger the nontrivial scalaron dynamics after the inflation.
Moreover, when the nonperturbative production is included, the e-folding $\tilde{N}$ could be also affected.
It is intriguing to study the possible scalaron oscillation around $\varphi=0$ on the effective potential and consecutive particle production and investigate how to connect to the radiation-dominant epoch.
\item 
The motion of SM particles (number densities) coupled to scalaron can be nonadiabatic when 
$\varphi \rightarrow + \infty$ 
due to the  exponential coupling form, $ \exp\left(-\sqrt{1/6}\kappa\varphi\right)$. 
This implies that nonperturbative production of the SM particles could happen  
during or immediately after the $R^2$ inflation, in the manner called the
{\it preheating}~\cite{Dolgov:1989us,Traschen:1990sw,Kofman:1994rk,Shtanov:1994ce,Kofman:1997yn}
(for reviews, see e.g., \cite{Kofman:1997yn,Amin:2014eta,Lozanov:2019jxc}).
This also tells us a possibility that the chameleon mechanism works even right after the inflation is over, 
hence the scalaron might not smoothly move down to the DE vacuum. 
This scenario might be similar to the situation known as the warm inflation~\cite{Berera:1995ie} instead of the standard cold inflation, 
the model parameters can thereby severely be constrained by cosmological observations. 
It is worth pursuing this {\it scalaron preheating} along with the SM particle production in detail.
\end{itemize}
At any rate,
the prospected scenarios above shall fill the missing link in the $F(R)$ gravity cosmology,
which deserves in other publications.

\begin{acknowledgments} 
We are grateful to Seishi Enomoto for the valuable discussion.
H.C. appreciates Taotao Qiu's insightful comments.
T.K. is supported by the National Key R\&D Program of China (2021YFA0718500).
S.M. is supported by the National Science Foundation of China (NSFC) under Grant No.11975108, 12047569, 12147217 and the Seeds Funding of Jilin University.
\end{acknowledgments}

\appendix

\section{Potential Analysis for Numerical Analysis}
\label{appendix A}

Despite the successful connection between inflation and DE, we face the hierarchy problem numerically.
The smallness of the DE scale suggests the gigantic separation from any other physical scale; however, we are interested in physics at such an intermediate scale.
For $F(R)$ model \eqref{F(R)_unification_model}, 
we examine the scalaron field $\varphi$ in Eq.~\eqref{Weyl transformation} and scalaron potential in Eq.~\eqref{Scalaron-vare-potential} in terms of Ricci scalar $R$:
\begin{align}
    e^{2\sqrt{1/6}\kappa\varphi} &= F^{\prime}_\text{DE}(R) + 2\alpha R 
    \\
    V\left(\varphi(R)\right) &= \frac{1}{2\kappa^{2}}
    \frac{ \left[RF^{\prime}_\text{DE}(R)-F_\text{DE}(R) \right] + \alpha R^{2}}
    {\left[ F^{\prime}_\text{DE}(R) + 2\alpha R \right]^{2}}
    \,.
\end{align}
The intermediate scale indicates $R_{c} \ll R$, and in Starobinsky's DE model
\begin{align}
    F_\text{DE}(R)=R -\beta R_{\text{c}}\left[1-\left(1+\frac{R^{2}}{R_{\text{c}}^{2}}\right)^{-n}\right]
    \, , \label{pure Starobinsky}
\end{align}
one finds
\begin{align}
    -\beta R_{\text{c}}\left[1-\left(1+\frac{R^{2}}{R_{\text{c}}^{2}}\right)^{-n}\right]
    \approx & -\beta R_{\text{c}}
    \,.
\end{align}
And thus, the total $F(R)$ function is approximated as GR with the cosmological constant and $R^{2}$ correction:
\begin{align}
    F(R) = R - \beta R_{\text{c}} + \alpha R^{2}
    \label{cc+inflation}
    \,.
\end{align}
The above is common even in other viable DE models of $F(R)$ gravity so that DE models of $F(R)$ gravity reproduce the $\Lambda$-CDM model in the large-curvature limit.
Thus, using the above $F(R)$ model for the large-curvature region $R_{c} \ll R$, 
we can investigate the $R^2$-corrected DE model or the unification of the inflation and DE in a model-independent way.

\begin{figure}[htbp]
    \centering
    \includegraphics[width=0.55\textwidth]{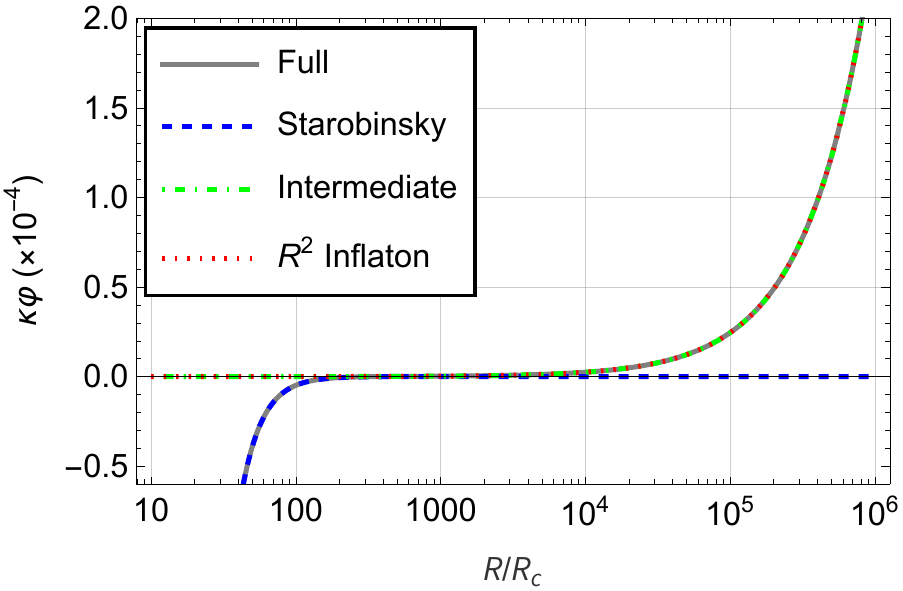}
    \caption{
    Comparison among the $R^2$-corrected Starobinsky model (gray solid) in Eq.~\eqref{Starobinsky DE}, 
    the Starobinsky model (blue dashed) in Eq.~\eqref{pure Starobinsky}, 
    approximated model for intermediate scale (green dot-dashed) in Eq.~\eqref{cc+inflation},
    and the $R^{2}$ inflation model (red dotted) with respect to the scalaron fields $\varphi$ as the function of the Ricci scalar $R$.
    The parameters are chosen as $n=1$, $\beta=2$, $\alpha R_{c}=10^{-10}$.
    }
    \label{Fig: field_comparison}
\end{figure}

\begin{figure}[htbp]
    \centering
    \includegraphics[width=0.55\textwidth]{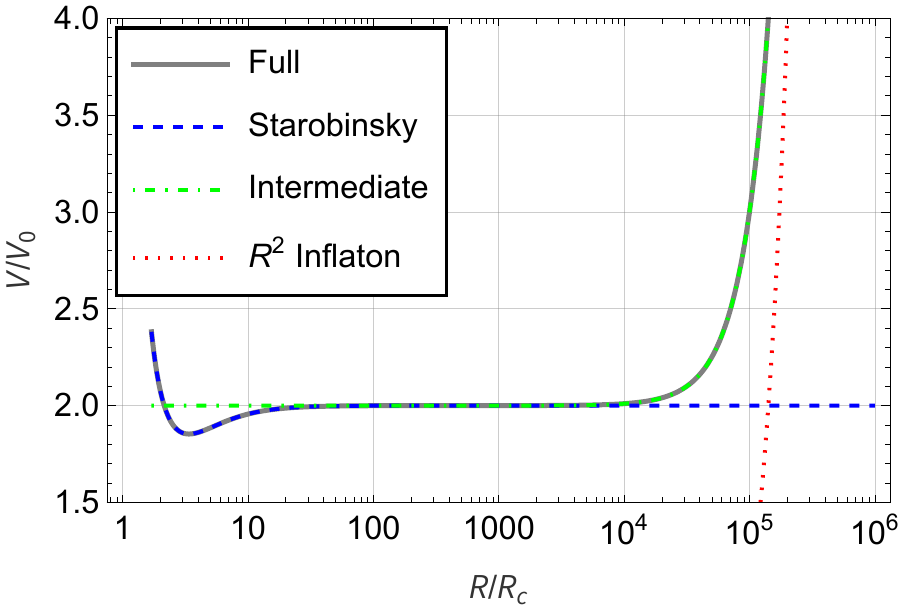}
    \caption{
    The same as Fig.~\ref{Fig: field_comparison} with respect to the scalaron potential $V(\varphi)$. 
    Eq.~\eqref{potential_approx} suggests that DE term and $R^2$ correction are comparable when $R/R_{c}=\sqrt{\beta/ \alpha R_{c}}\sim 10^{-5}$ in this setup.
        }
    \label{Fig: potential_comparison}
\end{figure}

For the model \eqref{cc+inflation}, the scalaron field and potential in terms of the Ricci scalar are given as 
\begin{align}
    e^{2\sqrt{1/6}\kappa\varphi} &= 1 + 2\alpha R 
    \,, \label{field_approx}
    \\
    V\left(\varphi(R)\right) 
    &= \frac{1}{2\kappa^{2}} \frac{\beta R_{\text{c}} + \alpha R^{2}}{\left[ 1 + 2\alpha R \right]^{2}}
    \,. \label{potential_approx}
\end{align}
Figs.~\ref{Fig: field_comparison} and \ref{Fig: potential_comparison} shows the scalaron field $\varphi$ and potential $V(\varphi)$ in terms of the Ricci scalar $R$. 
One can find that the approximated model for the intermediate scale is consistent with the full $R^{2}$-corrected one for the range $\varphi \geq 0$,
and it is not valid for the range $\varphi<0$ (see Fig.~\ref{Fig: bare_potential}).
Note that $\varphi \rightarrow 0$ and $V(\varphi) \rightarrow \text{const.}$ when $R\rightarrow \infty$, 
which shows the singularity problem~\cite{Frolov:2008uf}.

Furthermore, one finds an analytic form of the scalaron potential in terms of the scalaron field,
\begin{align}
    V(\varphi) 
    &= \frac{\beta R_{\text{c}} }{2\kappa^{2}} e^{-4\sqrt{1/6}\kappa\varphi} 
    + \frac{1}{8\kappa^{2}\alpha} \left(1 - e^{-2\sqrt{1/6}\kappa\varphi} \right)^{2}
    \label{scalaron_potential_cc}
    \,.
\end{align}
Note that the second term in the potential $V(\phi)$ represents the $R^2$ inflation potential.
Thus, the potential for $\varphi \geq 0$ is approximately described by Eq.~\eqref{scalaron_potential_cc} (see, Fig.~\ref{Fig: bare_potential}).
Although the potential for $\varphi<0$ should be written by the Starobinsky model~\eqref{pure Starobinsky},
the matter contribution $T^{\mu}_{\ \mu}$ in the effective potential Eq.~\eqref{effective potential} overwhelms the bare potential because of the hierarchy between the DE scale and the other physical scale.
Finally, one can conclude that in the existence of matter and chameleon mechanism, 
Eq.~\eqref{cc+inflation} and the trace of the energy-momentum tensor determines the scalaron dynamics in the $R^{2}$-corrected DE models of $F(R)$ gravity.

\section{Construction of kick function}
\label{appendix B}

In this section, we show how to handle data from Refs.~\cite{Husdal:2016haj, Bali:2014kia}. 
Data regarding the degrees of freedom for energy density $\rho$ and entropy density $s$ together with temperature have been listed in Table A1 in Ref.~\cite{Husdal:2016haj}. 
In order to study the effect of QCD PT, however, we also use the result on the equation of state from the lattice QCD simulation~\cite{Bali:2014kia}, ranging from $T=114-305$~[MeV], where data regarding entropy density $s$, energy density $\rho$, and pressure $P$ of quarks and mesons plasma have been provided~\footnote{
In the literature~\cite{Bali:2014kia} 
the lattice simulation has also incorporated 
nonzero magnetic fields so that the system gets 
anisotropic, hence the pressure $P$ is actually measured along the $z$-direction, $P_z$.     
The data plots that we quote in the present study,
include the cases with and without the applied magnetic field, 
from which we have taken only the case with zero magnetic field.  
}. 
We extract the degrees of freedom for entropy density from
\begin{align}
    g_{*s}(T)\equiv\frac{45}{2\pi^{2}}\frac{s(T)}{T^{3}}
    \,,
\end{align}
Furthermore, we have assumed that $g_{*}=g_{*s}$ during QCD PT, whose differences only appear after electron-positron annihilation.

Then, we set $T=305$~[MeV] and $T=114$~[MeV] as dividing points to separately deal with quantities before, in, and after QCD PT.
In particular, we add $14.25$, the remaining degree of freedom after the annihilation of $\pi^{0}$, into that during QCD PT by hand.
With the degree of freedom for the energy density in hand,
one obtains the energy density in Eq.~\eqref{energy density} and thereby obtains the kick function from Eq.~\eqref{kick function}.
The tabulated data regarding Fig.~\ref{Fig: dof} is available in \verb|degrees_of_freedom.dat| on arXiv.

\bibliographystyle{apsrev4-1}
\bibliography{references}

\end{document}